\documentclass[aps,prb,reprint,superscriptaddress]{revtex4-1}

\usepackage{hyperref}

\usepackage{graphicx}

\usepackage{amsmath}
\usepackage{bm}

\newcommand{\g}{|\vec H_\mathrm{iso}|}
\newcommand{\gl}{|\vec H_\mathrm{aniso}|}

\newcommand{\iso}{\vec H_\mathrm{iso}}
\newcommand{\aniso}{\vec H_\mathrm{aniso}}

\begin{document}

\title{Topological linelike bound states in the continuum}
\author{Manabu Takeichi}
\affiliation{Department of Physics, Tokyo Institute of Technology, 2-12-1 Ookayama, Meguro-ku, Tokyo 152-8551, Japan}
\author{Shuichi Murakami}
\affiliation{Department of Physics, Tokyo Institute of Technology, 2-12-1 Ookayama, Meguro-ku, Tokyo 152-8551, Japan}
\affiliation{TIES, Tokyo Institute of Technology, 2-12-1 Ookayama, Meguro-ku, Tokyo 152-8551, Japan}
\date{\today}

\begin{abstract}
Bound states in the continuum (BIC) have been studied  mainly in optics. Recently, electronic BIC have been proposed. They appear as points in the momentum space and are protected topologically by the Chern number. In this study, we propose a type of BIC protected by the winding number, which is  one of the topological invariants. These BIC appear as lines in the momentum space, and are realized in a multilayer model consisting of honeycomb-lattice layers. We show band structure and spatial localization of the BIC in this model. The wave numbers at which the BIC appear can be explained in terms of topology in the momentum space.
\end{abstract}
\maketitle

\section{introduction}
In a perfect crystal, eigenstates are Bloch states, which extend over the entire crystal. When the translational symmetry is locally broken, bound states may emerge. This can be seen in surfaces and interfaces, and around impurities and disorders. An example of bound states is  an impurity level in a semiconductor. Usually such bound states are perfectly  confined and exist outside the continuum in the band structure. However, bound states  can exist inside the continuum in some cases. These states are called bound states in the continuum (BIC)  [\onlinecite{bic}]. 
They have been proposed in many different fields such as photonics [\onlinecite{nature.541.196, PhysRevLett.100.183902, PhysRevLett.107.183901, PhysRevLett.111.240403, PhysRevLett.112.213903, PhysRevE.71.026611, PhysRevB.78.075105, PhysRevLett.113.037401, Longhi:14, doi:10.1021/acsphotonics.6b00860, natphoto.11.232, PhysRevLett.109.067401, Dreisow:09, PhysRevLett.111.220403}], acoustics [\onlinecite{PhysRevLett.118.166803, lyapina_maksimov_pilipchuk_sadreev_2015, LINTON200716}], and electronics [\onlinecite{scientific.8.5160, PhysRevB.85.115307, PhysRevB.73.205303, PhysRevB.73.235342, SABLIKOV20151775, PhysRevB.90.035434, PhysRevB.67.195335, PhysRevB.70.233315, 0295-5075-91-6-66001, PhysRevB.92.045409, PhysRevLett.109.116405, 0953-8984-25-23-235601, PhysRevB.89.115118}].

BIC have been studied mainly in optics. Optical BIC can be realized in a photonic crystal slab. Photonic band structure in a photonic crystal [\onlinecite{Joannopoulos:2008:PCM:1628775}] is divided  into two regions by the light line. Below the light line, the light can be perfectly confined within the slab because of the total reflection. On the other hand, above the light line, the light usually couples to the continuum and has a finite lifetime in general, leading to disappearance of bound states. Nevertheless, in some special photonic crystals, bound states appear above the light line, and they are called optical BIC [\onlinecite{nature499}]. In the wave vector space these BIC in Ref. [\onlinecite{nature499}] are located at vortex centers for the polarization directions of the far-field radiation.  They are protected by the winding number in the wave vector space, which means how many times the polarization vector winds around the wavevector of the BIC [\onlinecite{PhysRevLett.113.257401}].

BIC can exist for electrons in solids. In a previous study [\onlinecite{yang2013topological}], topologically protected electronic BIC have been proposed in a system of a two-dimensional (2D) quantum Hall insulator (QHI) [\onlinecite{PhysRevLett.61.2015}] stacked onto a  three-dimensional (3D) normal insulator (NI).  In this system, some of the electronic states from the 2D QHI can be degenerate  with the bulk bands formed by electronic states from the 3D NI. In general, states from the top QHI layer hybridize with other states in the 3D bulk because these states are not  orthogonal. On the other hand, BIC, originating from the 2D QHI, do not hybridize with other states from the 3D NI because these states are orthogonal. These BIC appear as points in the momentum space and the number of the point-like BIC is equal to the difference of the Chern number [\onlinecite{PhysRevLett.49.405}] between the  QHI and the NI, and therefore these BIC are topologically protected. 

In this study, we find a different type of BIC: line-like BIC in the 2D momentum space. The line-like BIC are topologically protected by the winding number. To demonstrate this, we use a model where a 2D isotropic honeycomb-lattice model is stacked on top of  a 3D anisotropic honeycomb-lattice model. The BIC derived from the 2D honeycomb-lattice layer do not hybridize with other states, and they localize near the topmost layer. These BIC emerge as lines in the momentum space, which is different from the previous work [\onlinecite{yang2013topological}]. This paper is organized as follows. In Sec \ref{two}, we first explain the honeycomb-lattice model.  In Sec \ref{three}, we show conditions for existence of BIC and  analytical results on the localization length of the BIC in this system. After that, we show band structures with and without BIC, and reveal that the BIC are protected by the winding number. Throughout the paper, we neglect electron-electron interaction and we consider spinless fermions.

\section{model for the line-like BIC\label{two}}

\begin{figure}[t]
\includegraphics[width=\columnwidth]{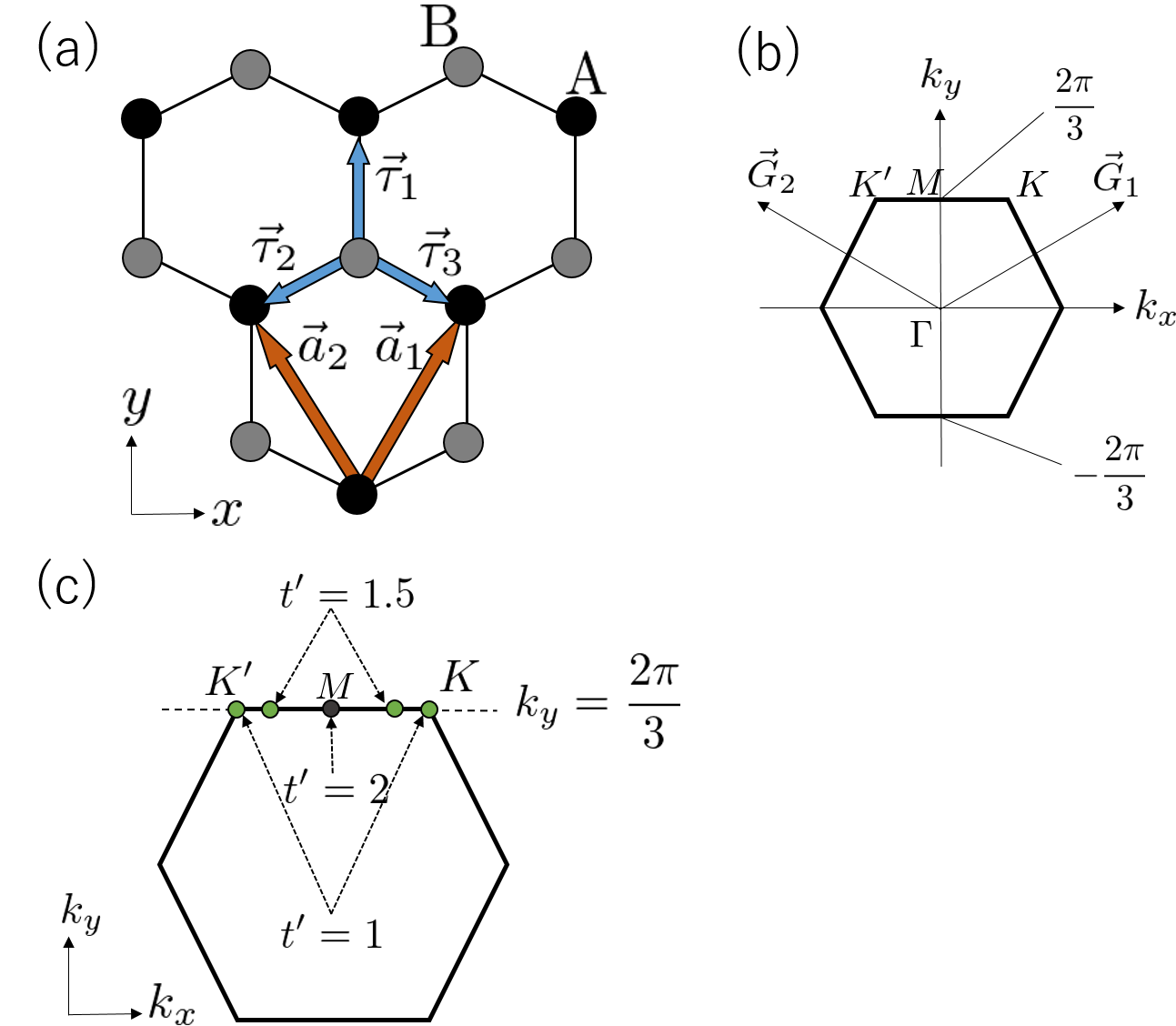}
\caption{(a) Honeycomb lattice used in the model. The honeycomb lattice has two sublattices labeled A and B. The red arrows  show primitive vectors, $\bm a_1=(\frac{\sqrt{3}}{2}, \frac{3}{2})$, $\bm a_2=(-\frac{\sqrt{3}}{2}, \frac{3}{2})$, and the blue arrows  represent vectors to three nearest neighbor sites, $\bm\tau_1=(0,1)$, $\bm\tau_2=(-\frac{\sqrt{3}}{2},-\frac{1}{2})$, $\bm\tau_3=(\frac{\sqrt{3}}{2},-\frac{1}{2})$. (b) First Brillouin zone of the honeycomb lattice. The reciprocal primitive vectors are $\bm G_1=2\pi\frac{2}{3}(\frac{\sqrt{3}}{2}, \frac{1}{2})$, $\bm G_2=2\pi\frac{2}{3}(\frac{-\sqrt{3}}{2}, \frac{1}{2})$. $\Gamma$, $M$, $K$, and $K'$ points are high-symmetry points in the Brillouin zone. (c) Dirac points for each layer of the 3D system in the momentum space at some values of $t'$. We show the Dirac points only along the line $k_y=\frac{2\pi}{3}$, as green symbols. At $t'=1$ the Dirac points are at $K$ and $K'$, as known in graphene. At $t'=2$ the two Dirac points meet and are annihilated.}
\label{schematic}
\end{figure}

When a 2D system is stacked on the top surface of a 3D system, electronic states from the 2D system can be degenerate with bulk states from the 3D system, and these states generally hybridize with each other. Therefore, the states derived from the 2D system are not localized near the surface. However, under some conditions, the states from the 2D system do not hybridize with bulk 3D states and localize near the surface. These states are called BIC.

\begin{figure}[t]
\includegraphics[width=\columnwidth]{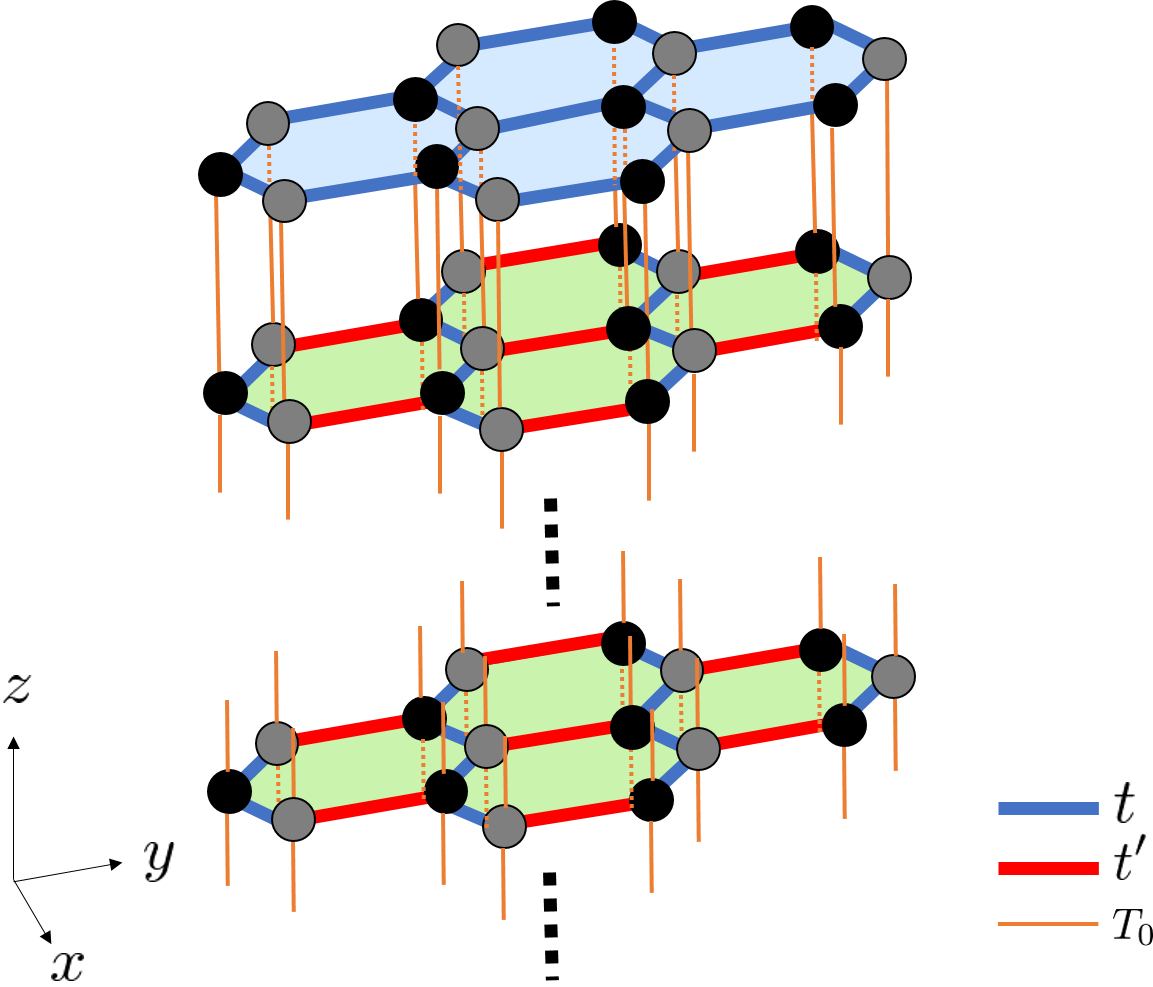}
\caption{Geometry of the system. It is a layered structure along the $z$ direction. The blue layer is the 2D system and the green layers constitute the 3D system.}
\label{stack}
\end{figure}

In this paper, we propose a  type of BIC, the line-like BIC in the momentum space. To show this we introduce a system of an isotropic 2D honeycomb-lattice model on top of a 3D system of a stacked anisotropic honeycomb-lattice model. The 2D system is an isotropic nearest-neighbor tight-binding model on the honeycomb lattice shown in Fig. \ref{schematic}(a), which is known to describe graphene [\onlinecite{RevModPhys.81.109}]. Each layer of the 3D system is similar to this 2D model, but with anisotropic hoppings to the nearest-neighbor sites [\onlinecite{PhysRevB.84.195452}]. We take the $xy$ plane to be along the layers, and the $z$-axis to be perpendicular to them. The stacking of the honeycomb lattice in this 3D system is along the $z$ axis as shown in Fig. \ref{stack}, unlike that in graphite. The Hamiltonian of the individual layer is described as
\begin{equation}
H=\sum_{\langle ij\rangle}t_{ij}c_i^{\dag}c_j,
\end{equation}
where $c_i$ is the annihilation operator of an electron at the site $i$, $t_{ij}$ is a real hopping amplitude, and $\langle ij\rangle$ denotes a pair of nearest-neighbor sites. We get the Bloch Hamiltonian,
\begin{align}
H(\bm k)=
	\begin{pmatrix}
		0 & \sum_{i=1}^3t_i\exp(-i\bm k\cdot\bm\tau_i) \\
		\sum_{i=1}^3t_i\exp(i\bm k\cdot\bm\tau_i) & 0 
	\end{pmatrix}, \label{hami}
\end{align}
where $\bm k$ is the Bloch wave vector, and $t_{i=1,2,3}$ are nearest-neighbor hopping amplitudes along the vectors $\bm\tau_{i=1,2,3}$ given by $\bm\tau_1=(0,1)$, $\bm\tau_2=(-\frac{\sqrt{3}}{2},-\frac{1}{2})$, and $\bm\tau_3=(\frac{\sqrt{3}}{2},-\frac{1}{2})$. Its energy eigenvalues are
\begin{align}
E(\bm k)=\pm\left|\sum_{i=1}^3 t_i\exp(i\bm k\cdot\bm\tau_i)\right|.
\end{align}
In our model (Fig. \ref{stack}) the three hopping parameters of the topmost 2D system are set to be equal, $t_{i=1,2,3}=t$, whereas in the underlying 3D system the hopping parameters are different, and we put $t_1=t'$ and $t_{i=2,3}=t$.  In this paper, we put $t=1$ for simplicity.

The Hamiltonian of the 2D isotropic system and that of the individual anisotropic layer of the 3D system are described as 
\begin{align}
H_{\mathrm{iso}}(k_x,k_y)
	=&\left( \cos k_y+2\cos\frac{\sqrt 3k_x}{2}\cos\frac{k_y}{2} \right)\sigma_x \notag \\
	&+\left( \sin k_y-2\cos\frac{\sqrt 3k_x}{2}\sin\frac{k_y}{2} \right)\sigma_y\\
	=&H_{\mathrm{iso},x}(k_x,k_y)\sigma_x+H_{\mathrm{iso},y}(k_x,k_y)\sigma_y \\
	=&\vec H_\mathrm{iso}(k_x,k_y)\cdot\vec\sigma ,
\end{align}
and
\begin{align}
H_{\mathrm{aniso}}(k_x,k_y)	
	=&\left( t'\cos k_y+2\cos\frac{\sqrt 3k_x}{2}\cos\frac{k_y}{2} \right)\sigma_x \notag \\
			&+\left( t'\sin k_y-2\cos\frac{\sqrt 3k_x}{2}\sin\frac{k_y}{2} \right)\sigma_y \\
=&H_{\mathrm{aniso},x}(k_x,k_y)\sigma_x+H_{\mathrm{aniso},y}(k_x,k_y)\sigma_y \\
=&\vec H_{\mathrm{aniso}}(k_x,k_y)\cdot\vec \sigma
\end{align}
respectively, where $\vec\sigma$ is the Pauli matrix acting onto the sublattice space. We call $\vec H_{\mathrm{iso}}=(H_{\mathrm{iso},x}, H_{\mathrm{iso},y}, 0)$ and  $\vec H_{\mathrm{aniso}}=(H_{\mathrm{aniso},x}, H_{\mathrm{aniso},y}, 0)$ pseudospins, and both of them lie along the $xy$-plane. The band structure of the 2D system has linear dispersion called Dirac cones near the $K$ and the $K'$ points in the momentum space shown in Fig. \ref{schematic}(b). The two degenerate points are called Dirac points. On the other hand, in the individual layers of the 3D system, the Dirac points are displaced from the $K$ and $K'$ points towards the $M$ point when the anisotropy is in the region $1<t'<2$, as shown in Fig. \ref{schematic}(c). Then the two Dirac points are annihilated in pair at $t'=2$ at the $M$ point. When $t'>2$, the band structure of each layer in the 3D system has no Dirac cones and has a gap.

Here, we note that the Dirac point is a vortex center of the pseudospin. It is because the Dirac point appears when the pseudospin vanishes, and around this Dirac point the pseudospin naturally acquires a vortex structure. One can associate each vortex with a vorticity which takes an integer value. The vorticity represents how many times the pseudospin rotates when we go around the vortex in a counterclockwise way. When $t'=1$ the Dirac points at $K$ and at $K'$ have vorticities $+1$ and $-1$, respectively. When $t'$ is increased from $t'=1$, the Dirac points in the individual layers of the 3D system move from the $K$ point to the $M$ point, and the vortex structure is modified.

We construct the 3D system  by stacking the layers described by $H_\mathrm{aniso}$ along  $z$ direction. The way of stacking is between A-A and B-B sites in Fig. \ref{stack}, and we put the interlayer hopping parameter $T_0$ to be  positive. Therefore, the  Bloch Hamiltonian of the 3D system is written as $H_{\text{3D-aniso}}(k_x,k_y,k_z)=H_{\mathrm{aniso}}+2T_0\sigma_0\cos k_z$ with the Bloch wave number $k_z$ along the $z$ direction and $\sigma_0$ being a $2\times 2$ unit matrix. Here we set the interlayer spacing to be unity.

\section{appearance of line-like BIC\label{three}}
\subsection{Conditions for appearance of BIC}
As is similar to the previous study [\onlinecite{yang2013topological}], we introduce the retarded Green's function for the purpose of describing the BIC. 
The effective Hamiltonian for this system can be considered as the Fano-Anderson model Hamiltonian [\onlinecite{PhysRev.124.1866}] and is written as
\begin{align}
H=&\sum_{k_x,k_y}\psi_{k_x,k_y}^\dagger(\iso\cdot\vec\sigma)\psi_{k_x,k_y} \notag \\
	&+ \sum_{k_x,k_y,k_z}\phi_{k_x,k_y,k_z}^\dagger(\aniso\cdot\vec\sigma+2T_0\cos k_z)\phi_{k_x,k_y,k_z} \notag \\
	&+ \sum_{k_x,k_y,k_z} \left(\psi_{k_x,k_y}^\dagger T_0\phi_{k_x,k_y,k_z}+\mathrm{H.c.}\right) ,
\end{align}
where $\psi_{k_x,k_y}$ represents the state in the 2D system and $\phi_{k_x,k_y,k_z}$ does the state in the 3D system. After integrating out the $\phi_{k_x,k_y,k_z}$, the thermal Green's function for the effective 2D system is written as
\begin{align}
G^{-1}=&i\omega_n-\iso\cdot\vec\sigma \notag \\
	&-T_0^2\int\frac{dk_z}{2\pi}\frac{i\omega_n-2T_0\cos k_z+\aniso\cdot\vec\sigma}{(i\omega_n-2T_0\cos k_z)^2-\aniso\cdot\aniso},
\end{align}
where $\omega_n$ is the Matsubara frequency.
The retarded Green's function $G_\mathrm{R}(\Omega)=G(i\omega_n=\Omega+i0^+)$ is described as
\begin{align}
G_{\mathrm R}^{-1}(\Omega) =\Omega -\vec H_\mathrm{iso}\cdot\vec\sigma +i\frac{\Gamma}{2}\left(1\pm \frac{\vec H_{\mathrm{aniso}}\cdot\vec\sigma}{\gl}\right). \label{green}
\end{align}
Here we performed the integral over $k_z$ by introducing the constant density of states $D_0$ for the 1D energy dispersion on $k_z$ direction.
In Eq. (\ref{green}), when $\Omega=2T_0\cos k_z+\gl$ ($\Omega=2T_0\cos k_z-\gl$) for some $k_z$, we have $\Gamma=\pi D_0T_0^2$ and take the positive sign (negative sign) for the coefficient of $\Gamma$, and for other values of $\Omega$, we have $\Gamma =0$. 
Bound states appear at the $\delta$-function singularity of $G_{\mathrm R}$, i.e., when $\mathrm{Det}(G_{\mathrm R}^{-1})=0$.
Because the condition of $\mathrm{Det}(G_{\mathrm R}^{-1})=0$ is rewritten as
\begin{align}
\mathrm{Det}(G_{\mathrm R}^{-1})=\Omega^2-\vec H_\mathrm{iso}^2+i\Gamma\left(\Omega\pm\frac{\vec H_{\mathrm{aniso}}}{\gl}\cdot\vec H_\mathrm{iso} \right)=0,
\end{align}
the electronic BIC emerge only in the two cases:
\begin{enumerate}
\item \hfill
\vspace{-\abovedisplayskip}
\vspace{-\baselineskip}
\begin{align}
\iso\cdot\aniso=\g\gl, \label{hei} \\
\Omega=2T_0\cos k_z\pm\gl =\mp\g \label{banhei}
\end{align}
or
\item \hfill
\vspace{-\abovedisplayskip}
\vspace{-\baselineskip}
\begin{align}
\iso\cdot\aniso=-\g\gl, \label{hanhei} \\
\Omega=2T_0\cos k_z\pm\gl =\pm\g. \label{banhanhei}
\end{align}
\end{enumerate}

The emergence of BIC is guaranteed by Eqs. (\ref{hei}) and (\ref{banhei}) or by Eqs. (\ref{hanhei}) and (\ref{banhanhei}). In either cases, there are two conditions  to realize BIC. Equations (\ref{hei}) and (\ref{hanhei}) means that the pseudospins $\iso$ and $\aniso$ are parallel and anti-parallel, respectively. Although they indicate  possible positions for emergence of the BIC, they are not sufficient to guarantee appearance of the BIC. Equation (\ref{banhei}) [Eq. (\ref{banhanhei})] is imposed when the pseudospins are parallel (anti-parallel).  Equations (\ref{banhei}) and (\ref{banhanhei}) guarantee that the states from the 2D system are embedded in the continuum bands. Then the states considered are BIC.

In the following, by calculating the band structure, we see that the BIC indeed appear when either of these conditions are satisfied.

\subsection{Analytical results\label{ana}}

To confirm the existence of  the BIC expected from the discussion in the previous subsection, we calculate the band structure for the present model in this subsection. The Hamiltonian of this system is written as
\begin{align}
H(k_x,k_y)=\left(
	\begin{array}{ccccc}
		           H_{\text{iso}}       &     T         &                  &                  &                    \\
		                T       & H_{\text{aniso}}    &      T         &                  &                    \\
		                         &    T          &   H_{\text{aniso}}    &     T           &                    \\
		                         &                &    T            &  H_{\text{aniso}}     &                    \\
		                         &                &                  &                   &   \ddots 　　 
	\end{array}
\right),   \label{5}
\end{align}
where $T$ is a $2\times 2$ matrix with $T_0$ in the diagonal components representing the hopping amplitude along the $z$ direction connecting the A-A and the B-B sites in Fig. \ref{stack}.

Now we analytically construct the BIC by truncating  wave functions  of the bulk Hamiltonian representing the 3D system.
The Schr$\ddot{\text{o}}$dinger equation of the bulk 3D system for the Bloch wave function $\phi_i(k_x,k_y)$ corresponding to the $i$-th eigen energy is written as
\begin{align}
\left(H_\mathrm{aniso}+2T_0\cos k_z\right)\phi_i(k_x,k_y)=E_i\phi_i(k_x,k_y),
\end{align}
where $\phi_i(k_x,k_y)$ is the periodic part of the $i$-th Bloch eigenfunction of the bulk 3D system, $k_z$ is the Bloch wave number along the $z$ axis, and $E_i$ is the corresponding energy. One can rewrite this equation into the following form:
\begin{align}
\left(
	\begin{array}{ccccc}
		           \ddots    &     　                  &                         &                     &                     		                  \\
		                         & H_{\text{aniso}}    &      T                &                      &                                              \\
		                         &    T                    &   H_{\text{aniso}}  &     T               &                                              \\
		                         &                         &    T                  &  H_{\text{aniso}}   &                                          \\
		                         &                        &                        &                      &    \ddots　 	        　 	
	\end{array}
\right)
	\begin{pmatrix}
		\vdots 	       	  \\
		\phi_{i}(k_x,k_y) \mathrm{e}^{0ik_z}    \\
		\phi_{i}(k_x,k_y) \mathrm{e}^{ik_z}    \\
		\phi_{i}(k_x,k_y) \mathrm{e}^{2ik_z}    \\
		\vdots  
	\end{pmatrix}      \nonumber    \\  
=E_i
	\begin{pmatrix}
		\vdots 			  \\
		\phi_{i}(k_x,k_y)   \mathrm{e}^{0ik_z}  \\
		\phi_{i}(k_x,k_y)   \mathrm{e}^{ik_z}  \\
		\phi_{i}(k_x,k_y)   \mathrm{e}^{2ik_z}  \\
		\vdots 
	\end{pmatrix}. \label{hbulk}
\end{align}
Let us assume that the eigenvector of the above equation, truncated  to a vector with the components with $\phi_i(k_x,k_y)\mathrm{e}^{ink_z}\ (n\ge 0)$, is an eigenvector of the Hamiltonian (\ref{5}) with the same energy $E_i$:
\begin{align}
\left(
	\begin{array}{cccc}
		           H_{\text{iso}}    & T    　                  &                         &                                     		                  \\
		             T            & H_{\text{aniso}}    &      T                &                                                                \\
		                         &    T                    &   H_{\text{aniso}}  &                                                                \\
		                         &                         &                      &  \ddots                      
	\end{array}
\right)
	\begin{pmatrix}
		\phi_{i}(k_x,k_y) \mathrm{e}^{0ik_z}    \\
		\phi_{i}(k_x,k_y) \mathrm{e}^{ik_z}    \\
		\phi_{i}(k_x,k_y) \mathrm{e}^{2ik_z}    \\
		\vdots  
	\end{pmatrix}      \nonumber    \\  
=E_i
	\begin{pmatrix}
		\phi_{i}(k_x,k_y)   \mathrm{e}^{0ik_z}  \\
		\phi_{i}(k_x,k_y)   \mathrm{e}^{ik_z}  \\
		\phi_{i}(k_x,k_y)   \mathrm{e}^{2ik_z}  \\
		\vdots 
	\end{pmatrix}. \label{hkei}
\end{align}
Equations (\ref{hbulk}) and (\ref{hkei}) are compatible with each other only when
\begin{align}
\left(H_{\text{iso}}+T_0\mathrm{e}^{ik_z}\right)\phi_i=E_i\phi_i,  \label{kahg}   \\
\left(H_{\text{aniso}}+2T_0\cos k_z\right)\phi_i=E_i\phi_i       \label{kahglike}
\end{align}
are both satisfied. To describe the bound states, we introduce an inverse of the localization length $\lambda(>0)$, and it is related with $k_z$ by $k_z=i\lambda$. 
We also consider the case with $\lambda\to\lambda + \pi i$, which stems from the chiral symmetry in the present system; if $\phi_i$ satisfies Eqs. (\ref{kahg}) and (\ref{kahglike}) with the energy $E_i$ and $k_z=i\lambda$, the state $\tilde\phi_i\equiv\sigma_z\phi_i$ also satisfies these equations with the energy $-E_i$ and $k_z=i(\lambda+\pi i)$.
Taking into account all of these, We can calculate the eigenstates and their localization lengths. Here we note that because of Eqs. (\ref{kahg}) and (\ref{kahglike}), $H_\mathrm{iso}$ and $H_\mathrm{aniso}$ commute. Then because $H_\mathrm{iso}$ and $H_\mathrm{aniso}$ are $2\times 2$ Hermitian matrices, the pseudospins $\vec H_\mathrm{iso}$ and $\vec H_\mathrm{aniso}$ should be parallel or anti-parallel. Through the calculations whose details are presented in Appendix \ref{ap}, we finally get the energy and the inverse of localization length
\begin{align}
	\lambda  & = \log \left|\frac{ \g -\gl }{T_0}\right|, \label{lam} \\
	E_{\mathrm{bound}}&=\pm\left(\frac{T_0^2}{\g-\gl} + \g\right)
\end{align}
when the pseudospins are parallel, and
\begin{align}
	\lambda  & = \log \frac{\g + \gl}{T_0}, \\
	E_\mathrm{bound}&=\pm\left(\frac{T_0^2}{\g+\gl} + \g\right) \label{ena}
\end{align}
when the pseudospins are anti-parallel. 

Let us apply these formulas to wavevector along the $K$-$M$-$K'$ line ($k_y=\frac{2\pi}{3}$, $-\frac{2\pi}{3\sqrt 3}\le k_x\le \frac{2\pi}{3\sqrt 3}$),
because the pseudospins are either parallel or anti-parallel  on this line. 
The inverse of localization length along the $k_y=\frac{2\pi}{3}$ line is written as
\begin{align}
\lambda =\log \frac{t'-1}{T_0} \label{eq:lambda}
\end{align}
for $t'>1$ because of $\g=-1+2\cos(\frac{\sqrt 3}{2}k_x)$ and $\gl=-t'+2\cos(\frac{\sqrt 3}{2}k_x)$. It implies that for given values of $t'$ and $T_0$, the localization length of the BIC  is constant along the $k_y=\frac{2\pi}{3}$ line if the BIC exist. In particular the BIC can exist for
\begin{align}
t'-1>T_0.
\end{align}
These results agree perfectly with those in Sec. \ref{num}.

We show how the inverse of the localization length Eq.~(\ref{eq:lambda}) behaves in various limits, and
explain that these behaviors match physically expectations. When $t'$ becomes smaller, the localization length diverges; it is reasonable since the BIC are expected to disappear as the
difference in hopping anisotropy between the surface and the bulk becomes smaller.  On the other hand, in the limit of interlayer hopping $T_0\rightarrow 0$, the localization length $\lambda^{-1}$ converges to zero, which is naturally expected. In the simultaneous limit of $t'\rightarrow 1$ and $T_0\rightarrow 0$ with $(t'-1)/T_0$ kept constant, $\lambda$ in
Eq. (\ref{eq:lambda}) remains a constant. Thus, the BIC seem to survive, which is unnatural because the difference in hopping anisotropy 
between the surface and the bulk becomes smaller. 
Nevertheless, it is not the case, and the BIC disappear in this limit, because the momentum space region with the BIC becomes narrower and vanishes in this limit. Thus, the behaviors of the inverse of the localization length in Eq.~(\ref{eq:lambda}) fit with physical expectations.

\subsection{Numerical results\label{num}}

\begin{figure*}[t]
\includegraphics[width=2\columnwidth]{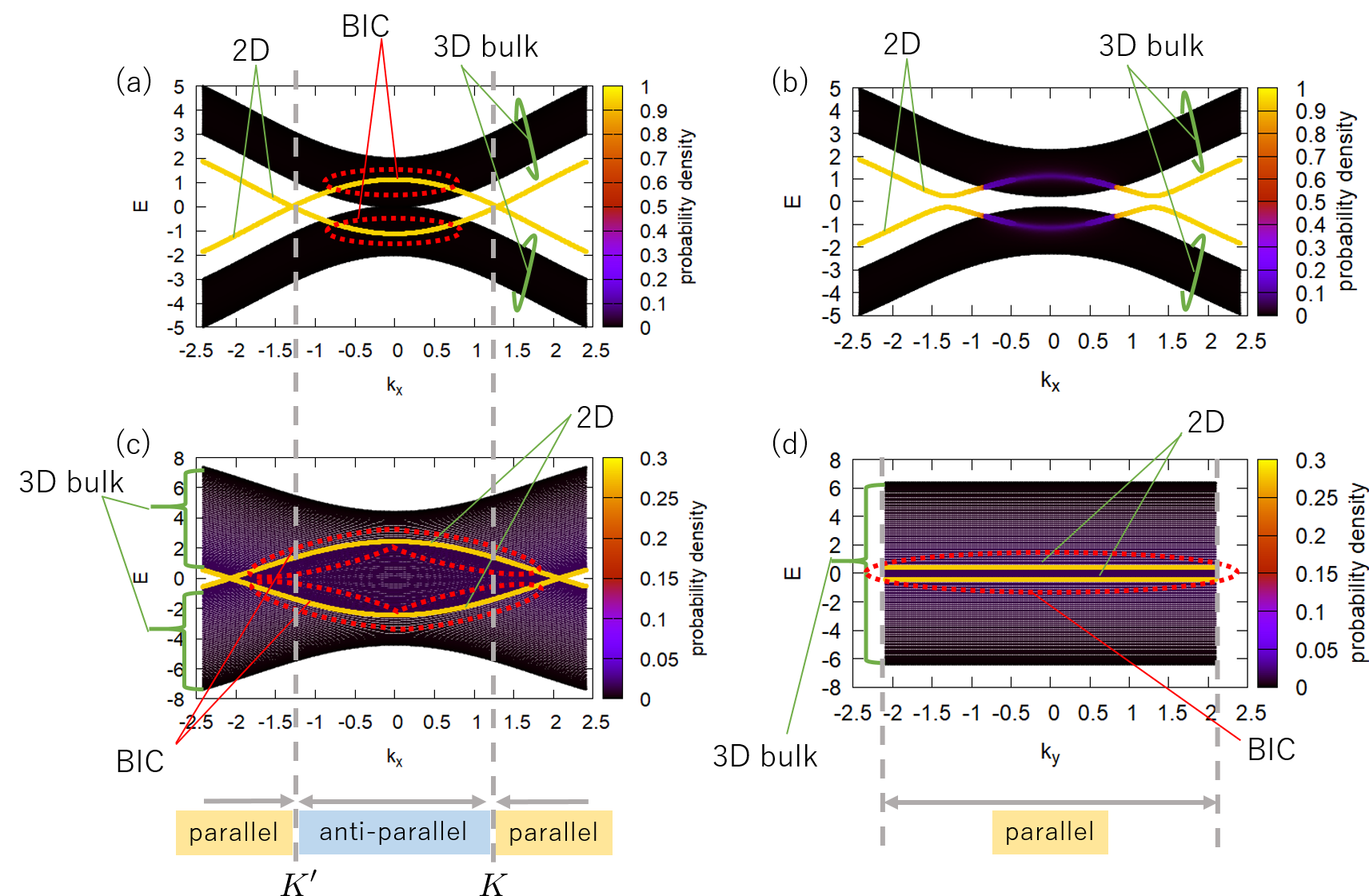}
\caption{Band structure for the present model consisting of the 2D system and the 3D anisotropic system. The colors represent the probability density at the top layer which is the isotropic 2D system. (a) Band structure along $k_y=\frac{2\pi}{3}$ for $t'=3$, and $T_0=0.5$. The yellow curves represent states from the topmost 2D system, and if they are embedded in the bands from the 3D system (shown in black), they are BIC. (b) Band structure along $k_y=\frac{3\pi}{5}$ for $t'=3$, and $T_0=0.5$. As the states from the topmost layer hybridize with bulk bands, there are no BIC. (c) Band structure along $k_y=\frac{2\pi}{3}$ for $t'=3$, and $T_0=1.7$. The BIC emerge in the wider region than that in (a). (d) Band structure along $k_x=\frac{\pi}{\sqrt 3}$ for $t'=3$, and $T_0=1.7$. The BIC are shown in yellow.
In (a), (c), and (d), the relative directions of the pseudospins $\iso$ and $\aniso$ are either parallel or anti-parallel as shown in the figure.
}
\label{koyutiprob}
\end{figure*}

Here we numerically calculate the band structure of the Hamiltonian (\ref{5}) along the $k_y=\frac{2\pi}{3}$ line as an example. Figure \ref{koyutiprob} (a) shows the band structure along the $k_y=\frac{2\pi}{3}$ line, with the probability density at the topmost 2D layer represented by the colors.  Here we set the anisotropy $t'=3$ and the hopping parameter along the $z$ direction $T_0=0.5$. In this figure, the broad bands shown in black are derived from the 3D system and the narrow band shown in yellow is derived from the 2D system. The probability density of the 2D states at the top layer is almost 1 throughout the entire Brillouin zone. Remarkably, this value remains almost unity, even in the region $-0.9\le k_x\le 0.9$ where these states become degenerate with those from the continuum band of the 3D system. Therefore, the states in this region derived from the 2D system are BICs and  are localized mostly at the top layer. 
On the other hand, when we focus on  $k_y=\frac{3\pi}{5}$ line instead of the $k_y=\frac{2\pi}{3}$ line, BIC do not emerge, as shown in Fig. \ref{koyutiprob}(b).
When the surface band enters the bulk bands from the 3D system at around $k_x\simeq \pm 0.9$, the probability density at the topmost layer becomes much lower.
Here the states from the topmost layer hybridize with the bulk bands from the 3D system.

\begin{figure*}[t]
\includegraphics[width=2\columnwidth]{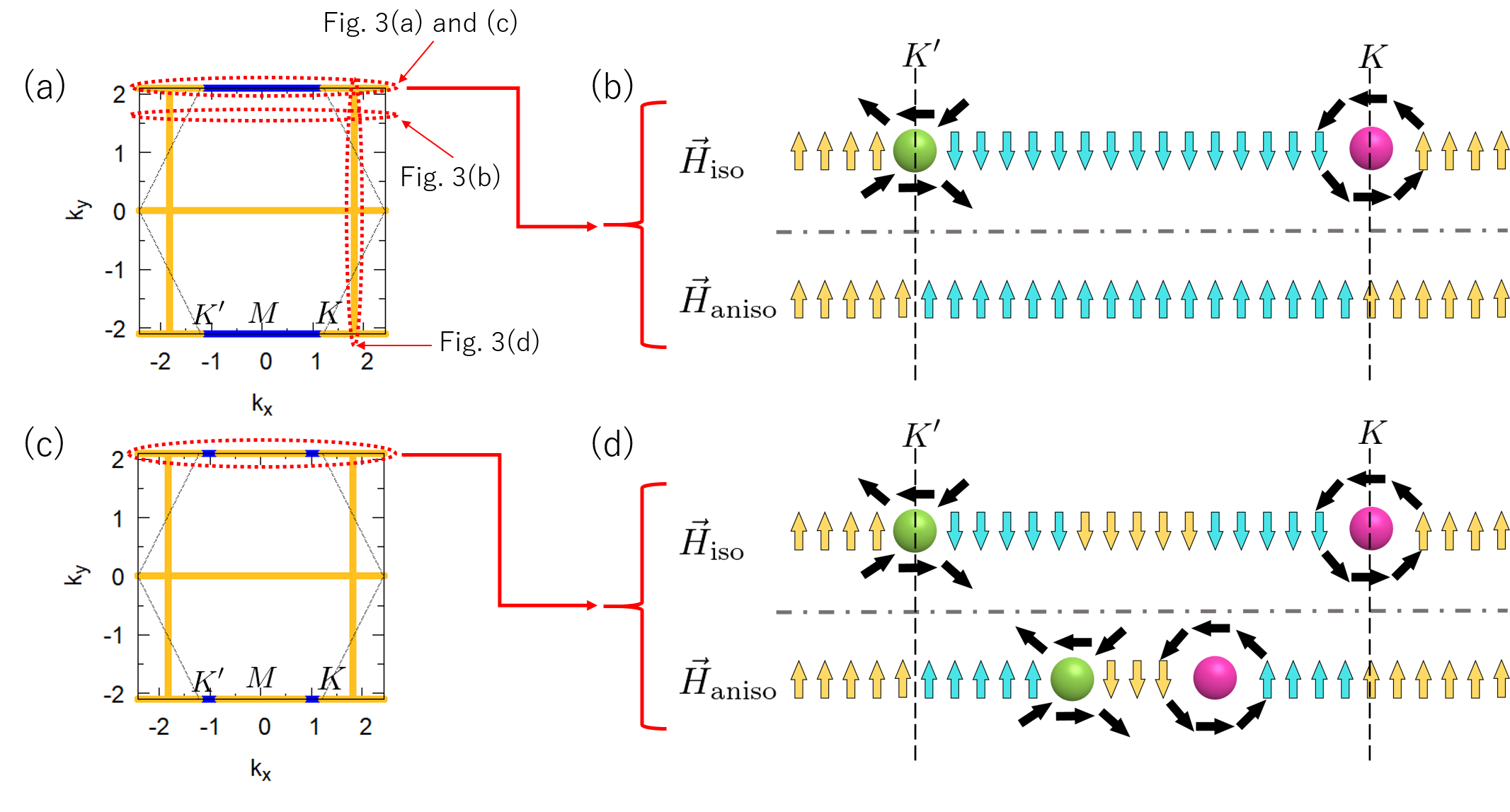}
\caption{ (a) and (c) Possible positions of the BIC in the momentum space  for (a) $t'=3$ and (c) $t'=1.5$. Yellow (blue) lines denote the cases when the pseudospins are parallel (anti-parallel). The hexagon represents the first Brillouin zone.
(b) and (d) Schematic diagram of the vortex structure along the $k_y=\frac{2\pi}{3}$ line for (b) $t'=3$ and (d) $t'=1.5$. The green and the red points represent Dirac points with opposite winding numbers, and the arrows mean the pseudospins on the $k_y=\frac{2\pi}{3}$ line. The upper panels in (b) and (d) represent the 2D system, and the lower ones show the individual layers of the 3D system. The BIC can exist only when the pseudospins $\iso$ and $\aniso$ are parallel (yellow arrows) or anti-parallel (blue arrows) in $k_y=\frac{2\pi}{3}$. Similar description holds true for $k_y=-\frac{2\pi}{3}, 0$ as well.
}
\label{bicmomentum}
\end{figure*}

This difference between the presence and absence of the BIC for $k_y=\frac{2\pi}{3}$ and $k_y=\frac{3\pi}{5}$, respectively, is understood from Eqs. (\ref{hei}) and (\ref{hanhei}). These equations indicate that only when the pseudospins $\iso$ and $\aniso$ are parallel or anti-parallel, the BIC can emerge.
Figure \ref{bicmomentum}(a) shows possible positions of BIC for $t'=3$ in the momentum space obtained from Eqs. (\ref{hei}) and (\ref{hanhei}). The yellow line means that 
$\iso$ and $\aniso$ are parallel, and the blue line means that they are anti-parallel. They imply that the states along $k_y=\pm\frac{2\pi}{3}, 0$ and $k_x=\pm\frac{\pi}{\sqrt 3}$  lines can become BIC.

Next we discuss how the size of the out-of-plane hopping $T_0$ affects the distribution of BIC.
First we discuss the case with $T_0=0.5$. Along the $k_y=\frac{2\pi}{3}$ line, among the states on the blue line in Fig. \ref{bicmomentum}(a) ($-\frac{2\pi}{3\sqrt 3}<k_x<\frac{2\pi}{3\sqrt 3}$), only those within the region $-0.9<k_x<0.9$ are BIC.
Meanwhile the states from the 2D system on the yellow line ($|k_x|>\frac{2\pi}{3\sqrt 3}$) in Fig. \ref{bicmomentum}(a) are not BIC as they do not enter the bulk bands. 
On the other hand, when  $T_0$ is relatively high ($T_0=1.7$) the states on the yellow line become BIC in Fig. \ref{koyutiprob}(c)  as can be understood from Eq. (\ref{banhei}). 
It is also the case along $k_x=\frac{\pi}{\sqrt 3}$ line [see Fig. \ref{koyutiprob}(d)].  As the hopping amplitude $T_0$ along the $z$ axis is relatively high in Figs. \ref{koyutiprob}(c) and \ref{koyutiprob}(d), the localization length of the states from the topmost layer is short.

The pseudospin texture changes when $t'$ is changed. Figure \ref{bicmomentum}(c) shows the wavevectors where the two pseudospins are parallel or anti-parallel for $t'=1.5$. At $t'=1.5$, the Dirac points for the individual layer of the 3D system lie on the $K$-$M$ and $K'$-$M$ lines. 
As we compare the cases with $t'=1.5$ [Fig. \ref{bicmomentum}(c)] and $t'=3$ [Fig. \ref{bicmomentum}(a)], we notice that the movement of the Dirac point due to the change of $t'$ flips the pseudospin direction of $\aniso$. As a result, along the $k_y=\frac{2\pi}{3}$ the pseudospins $\iso$ and $\aniso$ are parallel or anti-parallel both at $t'=1.5$ [Fig. \ref{bicmomentum}(d)] and $t'=3$ [Fig. \ref{bicmomentum}(b)], whereas their relative directions are different between $t'=1.5$ and $t'=3$ due to the change in the locations of Dirac points.

\subsection{Topological origin of the line-like BIC}
The appearance of the lines   $k_y=\pm\frac{2\pi}{3}, 0$ in Fig. \ref{bicmomentum}(a) can be understood from the topological nature of the pseudospin texture.
As the 2D isotropic system has Dirac points, the pseudospin texture forms vortices at these points. On the other hand, the individual layer of the 3D system does not have Dirac points when $t'>2$, and the pseudospin texture does not have vortices. Therefore along a loop around one of the Dirac points in the 2D system, $\iso$ rotates, whereas $\aniso$ does not, and therefore, there should be a point where $\iso$ and $\aniso$ are parallel and another point where they are anti-parallel. As  we  increase the size of the loop, these points form a line. In the present case this line is $k_y=\frac{2\pi}{3}$.
Along this line, $\iso$ and $\aniso$ are always parallel or anti-parallel as seen in Fig. \ref{bicmomentum}(b). 
Figure \ref{bicmomentum}(b) shows the vortex structures along the $k_y=\frac{2\pi}{3}$ for the 2D system on the upper side of this figure and that for the individual layer of the 3D system on the lower side of this figure. The green and the red points describe Dirac points, and the arrows mean pseudospins.
Therefore the BIC can exist along $k_y=\frac{2\pi}{3}$ from a topological origin and their existence is protected by the winding number. 
By changing the value of $t'$, the pseudospin structure of $\aniso$ changes. Meanwhile the topological nature of the Dirac points of $\iso$ at $K$ and $K'$ persists, and the yellow and the blue line remains along the $k_y=\frac{2\pi}{3}$ line.

Such line-like BIC have not been proposed previously, to the authors' knowledge. For example, the electronic BIC proposed in Ref. [\onlinecite{scientific.8.5160}] are extended in the momentum space. Nevertheless, unlike our paper the BIC in Ref. [\onlinecite{scientific.8.5160}] extend over the whole Brillouin zone because of the special symmetry in the lattice structure. Thus both the momentum space distribution and the physical origin of the BIC are quite different between our paper and Ref. [\onlinecite{scientific.8.5160}].

\section{conclusion}
In conclusion, we demonstrate existence of line-like BICs in the momentum space in the multilayer model which consists of  the isotropic 2D honeycomb-lattice system  stacked on top of the anisotropic 3D honeycomb-lattice system. When the states from 2D system are embedded in the continuum bands from the 3D system, if the states satisfy  specific conditions, they do not hybridize with states from the 3D system, and localize near the topmost 2D layer. 
These conditions mean that the states from the 2D system are orthogonal to the states from the 3D system.
These BIC are protected by the winding number of the pseudospin vector, and emerge as lines in the momentum space. 
This is in contrast with the previous study [\onlinecite{yang2013topological}], where the BIC emerge as points  protected by the Chern number in the momentum space. The difference of the momentum space distribution of the BIC stems from  the difference of the topological invariants that protect the BIC.

In the present paper, we studied a model with the isotropic 2D system stacked on top of the anisotropic 3D system. This is just an
example to demonstrate the line-like BIC. The key ingredient in this model is a difference in anisotropy between the top 2D layer and the bulk 3D system. This  difference gives rise to difference in the vortex distribution in the momentum space, leading to the line-like BIC from the topological
origin. Thus the line-like BIC exist in other models, for example in a model where the bulk 3D system is isotropic and the 2D system on the top layer is anisotropic.  Such an anisotropy difference between the surface and the bulk can be realized by buckling of the surface atomic layer.

\begin{acknowledgments}
We appreciate N. Nagaosa and B.-J. Yang for fruitful discussions. This work is supported by Grant-in-Aid for Scientific Research from MEXT (Grant No. 18H03678), and CREST, JST (No. JP-MJCR14F1).
\end{acknowledgments}

\appendix
\section{details of calculations on the energy and the localization length of BIC \label{ap}}
To calculate the properties of the BIC analytically in Sec. \ref{ana}, we start with Eqs. (\ref{kahg}) and (\ref{kahglike}). Because these equations must have the same eigenvalue and eigenstate, there are two distinct cases:

(I) 
\begin{align}
&E_i=T_0\mathrm e^{ik_z}\pm\g=2T_0\cos k_z\pm\gl, \\
&\phi_i=
	\begin{pmatrix}
		1 \\
		\pm\frac{H_{\mathrm{iso},x}+iH_{\mathrm{iso},y}}{\g}
	\end{pmatrix}
	=
	\begin{pmatrix}
		1 \\
		\pm\frac{H_{\mathrm{aniso},x}+iH_{\mathrm{aniso},y}}{\g}
	\end{pmatrix}
\end{align}
meaning that $\vec H_\mathrm{iso}$ and  $\vec H_\mathrm{aniso}$ are parallel, and

(I\hspace{-.1em}I)
\begin{align}
&E_i=T_0\mathrm e^{ik_z}\pm\g=2T_0\cos k_z\mp\gl, \\
&\phi_i=
	\begin{pmatrix}
		1 \\
		\pm\frac{H_{\mathrm{iso},x}+iH_{\mathrm{iso},y}}{\g}
	\end{pmatrix}
	=
	\begin{pmatrix}
		1 \\
		\mp\frac{H_{\mathrm{aniso},x}+iH_{\mathrm{aniso},y}}{\g}
	\end{pmatrix}
\end{align}
meaning that $\vec H_\mathrm{iso}$ and  $\vec H_\mathrm{aniso}$ are anti-parallel.

We discuss these two cases (I) and (I\hspace{-.1em}I) separately.
First, we study the case (I).
To describe the bound states, we put $k_z=i\lambda\ (\lambda>0)$, and get the energy of the bound state $E_\mathrm{bound}$
\begin{align}
&E_\mathrm{bound}=T_0\mathrm e^{-\lambda}\pm\g=2T_0\cosh\lambda\pm\gl, \\
&\mathrm e^\lambda=\pm\frac{\g-\gl}{T_0}.
\end{align}
We also consider the case with replacement $\lambda\to\lambda+\pi i$:
\begin{align}
&E_\mathrm{bound}=-T_0\mathrm e^{-\lambda}\pm\g=-2T_0\cosh\lambda\pm\gl, \\
&\mathrm e^\lambda=\mp\frac{\g-\gl}{T_0}.
\end{align}
By noting $\mathrm e^\lambda>0$, these are rewritten as follows. When $\g>\gl$,
\begin{align}
&E_\mathrm{bound}=\pm\left(T_0\mathrm e^{-\lambda}+\g\right)=\pm\left(2T_0\cosh\lambda+\gl\right), \\
&\lambda=\log\frac{\g-\gl}{T_0}
\end{align}
or when $\g<\gl$,
\begin{align}
&E_\mathrm{bound}=\pm\left(T_0\mathrm e^{-\lambda}-\g\right)=\pm\left(2T_0\cosh\lambda-\gl\right), \\
&\lambda=\log\frac{-\g+\gl}{T_0}.
\end{align}

Next, we consider the case (I\hspace{-.1em}I) in the same way. We put  $k_z=i\lambda\ (\lambda>0)$ to describe the bound states, and get
\begin{align}
&E_\mathrm{bound}=T_0\mathrm e^{-\lambda}\pm\g=2T_0\cosh\lambda\mp\gl, \\
&\mathrm e^\lambda=\pm\frac{\g+\gl}{T_0}.
\end{align}
We also consider the case with replacement $\lambda\to\lambda+\pi i$:
\begin{align}
&E_\mathrm{bound}=-T_0\mathrm e^{-\lambda}\pm\g=-2T_0\cosh\lambda\mp\gl, \\
&\mathrm e^\lambda=\mp\frac{\g+\gl}{T_0}.
\end{align}
They are rewritten as
\begin{align}
&E_\mathrm{bound}=\pm\left(T_0\mathrm e^{-\lambda}+\g\right)=\pm\left(2T_0\cosh\lambda-\gl\right), \\
&\lambda=\log\frac{\g+\gl}{T_0}.
\end{align}
Thus we get the results in Eqs. (\ref{lam})-(\ref{ena}).


%

\end{document}